\documentclass{article}
\usepackage[top=1.5in, bottom=1.5in, left=1.2in, right=1.2in]{geometry}
\usepackage{authblk}
\usepackage{amssymb, amsmath,framed}
\usepackage{graphicx}
\usepackage[round]{natbib}

\usepackage{bm}
\usepackage[colorlinks=true,linkcolor=blue,urlcolor=blue,citecolor=blue,anchorcolor=blue]{hyperref}
\expandafter\ifx\csname package@font\endcsname\relax\else
 \expandafter\expandafter
 \expandafter\usepackage
 \expandafter\expandafter
 \expandafter{\csname package@font\endcsname}
\fi
\def\bq{\begin{equation}}
\def\eq{\end{equation}}
\def\bqy{\begin{eqnarray}}
\def\eqy{\end{eqnarray}}

\makeatother

\begin{document}

\title{Relative Likelihood of Success in the Searches for Primitive versus Intelligent Extraterrestrial Life}

\author{Manasvi Lingam\thanks{Electronic address: \texttt{manasvi.lingam@cfa.harvard.edu}}\,\,}

\author{Abraham Loeb\thanks{Electronic address: \texttt{aloeb@cfa.harvard.edu}}}
\affil{Institute for Theory and Computation, Harvard University, 60 Garden St, Cambridge MA 02138, USA}

\date{}

\maketitle

\begin{abstract}
We estimate the relative likelihood of success in the searches for primitive versus intelligent life on other planets. Taking into account the larger search volume for detectable artificial electromagnetic signals, we conclude that both searches should be performed concurrently, albeit with significantly more funding dedicated to primitive life. 
Based on the current federal funding allocated to the search for biosignatures, our analysis suggests that the search for extraterrestrial intelligence (SETI) may merit a federal funding level of at least $\$10$ million per year, assuming that the average lifetime of technological species exceeds a millennium.
\end{abstract}

\section{Introduction} \label{SecIntro}
The major advances in exoplanetary science over the past few years are well documented. After the \emph{Kepler} mission was launched in 2009 \citep{Bat14,Bor16},\footnote{\url{https://www.nasa.gov/mission_pages/kepler/overview/index.html}} the number of detected exoplanets is in the thousands. This field has received a major boost within the last couple of years owing to two exciting discoveries. The first was the discovery of a terrestrial planet in the habitable zone (HZ) of Proxima Centauri, which is the star nearest to the Sun \citep{AE16}. The second was the discovery of seven Earth-sized planets transiting the ultracool dwarf star TRAPPIST-1 at a distance of $12$ pc, of which three may lie within the HZ \citep{GJ16,GT17}. The detection of a temperate super-Earth orbiting the cool star LHS 1140 located $12$ pc from the Sun also merits a mention in this context \citep{DIC17}. The search for life on exoplanets is expected to receive a major boost with the launch of the James Webb Space Telescope (JWST),\footnote{\url{https://www.jwst.nasa.gov/}} planned for launch in 2021, which will enable the characterization of exoplanetary atmospheres and searching for biologically produced gases \citep{FA18,SKP18}. For instance, theoretical models indicate that the spectral atmospheric features for six of the TRAPPIST-1 planets can be detected in fewer than 20 transits \citep{MK17}.

While the search for biosignatures is primarily oriented towards confirming the existence of ``primitive'' life, the detection of technosignatures would indicate the presence of ``intelligent'' life.\footnote{By ``primitive'' life, we refer to microbes for the most part, although complex (but non-technological) multicellular life also falls under this category. Similarly, ``intelligent'' life refers to technologically advanced intelligence, i.e. to species that are capable of producing detectable signatures of their technology. It must be noted that ``intelligent'' species will not always possess advanced technology, as seen from the examples of cetaceans on Earth.} In reality, it must be observed that this classification is somewhat facile because certain biosignatures could end up being conflated with technosignatures. An interesting example in this context was pointed out by \citet{Raup}, who suggested that certain species can naturally evolve communication in the radio frequencies without necessarily being technologically advanced and may therefore be mistaken for signatures of technological intelligence. Hence, the possibility that the boundaries between biosignatures and technosignatures could become blurred must be borne in mind. 

The search for technosignatures resides in the domain of the Search for Extraterrestrial Intelligence (SETI), which has had a chequered history. The pioneering work by \citet{CM59}, followed by Project Ozma and the introduction of the well-known Drake equation \citep{Dra61}, facilitated the rapid rise of SETI in the 1960s. While SETI remained funded to varying degrees during the 1970s and 1980s, a multitude of sociocultural factors led to the cancellation of NASA's SETI program in 1993 \citep{Gar99,WOR18}. The next two decades proved to be difficult for SETI research as it had to survive with relatively limited private funding. In recent times, the funding for SETI has witnessed a distinct upswing in fortunes. This is primarily due to the inauguration of the Breakthrough Listen initiative \citep{WD17,IS17},\footnote{\url{https://breakthroughinitiatives.org/initiative/1}} the largest SETI undertaking to date that allocates a total of $\$100$ million over the next decade. Furthermore, an intriguing bill is also under consideration in the USA House of Representatives that proposes to allocate a total of $\$10$ million over a two-year timespan to search for technosignatures.\footnote{\url{https://democrats-science.house.gov/sites/democrats.science.house.gov/files/documents/NASA2018_002_xml.pdf}}

In summary, the immediate future for detecting biosignatures and technosignatures appears to be bright. In this paper, we seek to evaluate the relative likelihood of successfully detecting primitive and intelligent extraterrestrial life using state-of-the-art telescopes. We focus primarily on extrasolar systems within the Milky Way, and exclude searches for life within our Solar system and outside our Galaxy in our analysis. By estimating the relative likelihood, we address the question of how much funding should be allocated to SETI per year.

\section{Estimating the likelihood of detecting life via technosignatures and biosignatures}
In order to estimate the number of worlds with life that can be found by means of biosignatures or technosignatures, we shall adopt the following approach. Broadly speaking, the number of worlds that can be detected can be expressed as the product of the number of worlds with life within a given survey volume,\footnote{By ``survey volume'', we refer to the maximum possible volume that can be spanned by a particular telescope for detecting signs of life (either biosignatures or technosignatures).} and the probability of detecting life by means of a particular method and state-of-the-art observatories. Hence, it is important at this stage to reiterate that our results depend on: (A) the search strategies employed, and (B) the current technological level. We shall briefly go over the ramifications arising from these two criteria in Secs. \ref{SSecBio} and \ref{SSecET}.

We consider two broad avenues for detecting life. The first is by means of searching for biosignatures, which enables the detection of simple/complex (but non-technological) life. The second is via searches for technosignatures, which enable the detection of technological (and ``intelligent'') life. In other words, we aim to compute the following quantities:
\begin{equation}\label{Nbd}
    N_b = \mathcal{N}_b \cdot \mathcal{P}_b,
\end{equation}
where $N_b$ is the number of worlds with life detectable by means of biosignatures, $\mathcal{N}_b$ is the number of worlds with life within a particular survey volume, and $\mathcal{P}_b$ is the probability of detecting life by means of biosignatures for this particular survey. In a similar vein, we have
\begin{equation}\label{Ntd}
      N_t = \mathcal{N}_t \cdot \mathcal{P}_t,
\end{equation}
with $\mathcal{N}_t$ denoting the number of worlds with technological species in the survey volume, and $\mathcal{P}_t$ is the probability of detecting technosignatures using a given search strategy. We are ultimately interested in the ratio $\delta$ defined as
\begin{equation}\label{delta}
    \delta = \frac{N_t}{N_b},
\end{equation}
since it quantifies the number of worlds with life that can be detected by means of technosignatures versus biosignatures; hence, we can treat it as the relative likelihood (RL) of detecting life. 

\subsection{Number of worlds detectable through biosignatures}\label{SSecBio}
We will now attempt to estimate $N_b$ by using a Drake-type equation. As noted earlier, this depends on two distinct criteria, namely (A) and (B). We shall adopt transmission spectroscopy as the method of detecting biosignatures and assume that the telescope under consideration is the JWST. In principle, high-contrast direct imaging could enable the detection of a wider range of biosignatures, but the requisite telescopes are expected to become operational in only about a decade from now \citep{Kal17,FA18}, such as the $30$ m class ground telescopes and potentially the space-based Wide Field Infrared Survey Telescope (WFIRST). Hence, we restrict ourselves our attention to detecting biosignature gases on exoplanets orbiting M-dwarfs by means of transit spectroscopy.

We begin by estimating $\mathcal{N}_b$, which can be expressed as the product,
\begin{equation}\label{Ncalb}
    \mathcal{N}_b = N_\mathrm{sur} \cdot f_e \cdot f_l,
\end{equation}
where $N_\mathrm{sur}$ is the number of stars that can be covered by a state-of-the-art telescope like the JWST, $f_e$ is the fraction of ``habitable'' planets per star, and $f_l$ is the probability that a ``habitable'' planet is actually inhabited. It is important to correct a common misconception: $f_e$ does not represent the fraction of stars hosting Earth-sized planets in the habitable zone, since this is not a necessary and sufficient condition for guaranteeing habitability \citep{SMG16}. As a result, $f_e$ remains an unknown quantity since we do not currently know the list of necessary and sufficient criteria for habitability. Similarly, we express $\mathcal{P}_b$ as 
\begin{equation}\label{Probb}
    \mathcal{P}_b = f_q \cdot f_t \cdot f_d,
\end{equation}
where $f_q$ is the fraction of stars that are non-flaring, $f_t$ is the fraction of planets that are transiting and observable by a telescope like the JWST, and $f_d$ is the probability of detecting biosignature gases provided that life does exist on a given planet. We focus only on non-flaring stars since active stars (with frequent flaring activity) can alter the concentrations of biosignature gases in the atmosphere and also pose difficulties for habitability \citep{Gren17}. It should, however, be appreciated that flares can result in enhanced levels of UV radiation and solar energetic particles that may be advantageous to the origin of life \citep{RXT18,LDF18}, but these factors could also pose impediments to the emergence of complex land-based life \citep{LiLo}.

Here, $f_d$ quantifies the fact that planets can give rise to \emph{both} false negatives and false positives insofar detecting life via biosignature gases is concerned. As an example, consider oxygen, which is a prominent biosignature gas. For about half of Earth's evolutionary history, the oxygen in the atmosphere was less than $1\%$ present atmospheric level (PAL) even when life was quite abundant, implying that the non-detection of life through this approach would have led to a false negative result \citep{Pil03,ROS17}. Similarly, there are several avenues by which abiotic O$_2$ can be produced in considerable abundance (e.g. photodissociation of H$_2$O and CO$_2$), thereby leading to potential false positives \citep{MRA18,CKT18}. Other biosignature gases with a lower false positive/negative probability can be considered, but their atmospheric concentrations are likely to be much lower and therefore harder to detect. 

When combined, we note that (\ref{Ncalb}) and (\ref{Probb}) are virtually identical to the `Biosignature Drake Equation' recently proposed by \citet{Sea17}.\footnote{\url{https://www.cfa.harvard.edu/events/2013/postkepler/Exoplanets_in_the_Post_Kepler_Era/Program_files/Seager.pdf}} If we suppose that planets up to a distance of $\sim 30$ pc ($\sim 100$ lt yrs) can be studied by the JWST, we get $N_\mathrm{sur} \sim 3 \times 10^4$ stars. We choose $f_q \sim 0.2$ and $f_t \sim 10^{-3}$ as these parameters are reasonably constrained, but it should be noted that $f_d$ remains unknown. In our analysis, we will retain this variable but normalize it by a conservative fiducial value of $0.1$. We have not provided values for $f_e$ and $f_l$ either, but it will be seen subsequently that they drop out from our analysis.

\subsection{Number of worlds detectable through technosignatures}\label{SSecET}
The strategies for detecting technosignatures are numerous. Broadly speaking, they can be divided into two (non-exclusive) categories: electromagnetic technosignatures (e.g. radio and optical signals) and artifact technosignatures (e.g. megastructures and artificial lights); for reviews of this subject, the reader may consult \citet{Tart01}, \citet{Wil01}, \citet{WM14} and \citet{Cab16}. The survey volume, i.e. the distance to which a particular technosignature can be detected through a given method, also varies considerably. We shall restrict ourselves only to Galactic SETI, and defer the discussion of extragalactic SETI \citep{WG14,ZCA15} to Sec. \ref{SecASETI}.

We will focus on the case where a ``conventional'' SETI search for radio or optical signals is being carried out.\footnote{It should be noted that electromagnetic signals may also be detectable due to the leakage radiation emitted as a consequence of powering light sails \citep{GL15,BB16,LiMa18} even at extragalactic distances \citep{Li17}.} We will furthermore assume that the signals are being broadcast deliberately, since it has been proposed that even the Square Kilometre Array (SKA) would have a very low probability of detecting accidental signal leakage from human-level technology \citep{FN11}. In contrast, we note that targeted signals broadcast by Arecibo could be detected at a distance of $\gtrsim 1$ kpc by a receiver of similar capabilities \citep{Tart01}. More powerful beacons lie within the capabilities of present-day human civilization and would cost a few billion USD \citep{BBB10}. Similarly, it has been argued that the detection threshold for continuous wave lasers is between a few kW and MW (assuming a 10 m sized telescope), which falls within the bounds of current or upcoming human technology \citep{TM17}. If optical pulses are used, even with present-day lasers, it is expected that they will outshine the visible light of the Sun by 4 orders of magnitude \citep{HH04}. Thus, for radio/optical SETI involving the deliberate broadcasting of artificial signals \citep{Bor12}, it seems reasonable to suppose that the survey volume can be taken to encompass most of the Galaxy if the technological level of the transmitting species is more advanced than, or comparable to, that of present-day humans. If the Solar gravitational lens were to be utilized in the future for receiving artificial electromagnetic signals, this will enhance the reach of SETI missions by many orders of magnitude \citep{Esh79,Hip18}.

For a Galactic SETI survey, the value of $\mathcal{N}_t$ is given by a variant of the famous Drake equation \citep{Dra61,SS66} as follows:
\begin{equation}\label{Ncalta}
    \mathcal{N}_t = \alpha \cdot N_g \cdot f_e \cdot f_l \cdot f_i \cdot \frac{L}{t_\star}.
\end{equation}
Note that $N_g \sim 10^{11}$ is the number of stars in our Galaxy, $f_i$ denotes the fraction of life-bearing planets that subsequently give rise to technological species with the capacity for taking part in detectable interstellar communication,\footnote{In the Drake equation, this is typically expressed as the product of two distinct factors \citep{SS66} which we have collapsed into one for the sake of brevity.} $L$ is the average lifetime of a technological species and $t_\star \sim 10^{10}$ yrs represents the typical lifetime of a solar-type star as well as the approximate age of the Milky Way. Note that the last factor $L/t_\star$ can be interpreted as the fraction of technological species that are \emph{currently} active in our Galaxy \citep{Mac11,FS16}. We have implicitly assumed that the number of suitable planets per star is around unity, and that all stars are equally suitable for hosting life; the latter is not necessarily valid since there are several factors which might make M-dwarfs inhospitable for life \citep{LL18,Man18,LiLo}.\footnote{It must, however, be appreciated that planets around M-dwarfs could still serve as outposts for technologically advanced life as these stars are extremely long-lived, thereby providing a ready energy source for up to trillions of years.} We have introduced the additional factor $\alpha$ to account for the fact that all advanced technological species need not be situated on planets. They can spread outwards and settle other worlds or even dwell in interstellar space; this may be particularly feasible if the technological species is post-biological. Thus, $\alpha$ can be regarded as an averaged ``settlement'' factor that is present in some versions of the Drake equation \citep{WHK80,Brin}, and it can become much greater than unity in principle \citep{Ling16}.\footnote{Another possible phenomenon that can raise the value of $\alpha$ is panspermia \citep{Arr08,Wick10}, which is expected to be particularly effective in certain astrophysical environments such as globular clusters \citep{DSR16}, the Galactic centre \citep{MG15,CFL18}, and M-dwarfs \citep{Ling17}.} 

We will henceforth employ the optimistic fiducial values of $0.1$ for $f_i$ and $\sim 10^4$ yr for $L$ \citep{VH61,Gott93,Chi15}; some formulations of the Drake equation have advocated that $f_i \sim 1$ is conceivable based on the alleged continuity between human and non-human minds \citep{Dar,Mar15}. However, we wish to emphasize the fact that our choices for both $f_i$ and $L$ are truly unknown and by no means definitive, and therefore alternative values can be utilized instead. There have been several proponents who have argued in favor of a very low value of $f_i$ on the grounds that technological intelligence has evolved only once on our planet, and that too rather late in our evolutionary history given that the Earth is expected to be habitable for only $1$-$2.5$ Gyr in the future \citep{RCO13}. If we choose, for example, $f_i \sim 10^{-8}$, we would end up with $\mathcal{N}_t \lesssim 1$ even for optimistic choices of the other parameters. The net effect of a low value of $f_i$ is that searches for intelligent life will be highly disfavored relative to searches for primitive life. We will further discuss the implications of choosing alternative values for $f_i$ and $L$ in Sec. \ref{SecDel}.

The probability of detection $\mathcal{P}_t$ may be written as
\begin{equation}\label{probt}
    \mathcal{P}_t = f_{cs} \cdot \varepsilon_d.
\end{equation}
The first factor ($f_{cs}$) denotes the fraction of current technological species that are broadcasting signals detectable by humans. If a species happens to be much more advanced than the Earth, then it may either be uninterested in sending signals to humans or rely upon channels that are not decipherable by us \citep{Sag73}. The second factor ($\varepsilon_d$) is the probability of intercepting a detectable signal, from a sufficiently advanced technological species, by the Earth. Another notable factor that we will not consider here is the identification of the frequencies being used by the transmitting species \citep{DS73} since it cannot be easily quantified. Hence, we assume that the transmitter and receiver are cognizant of the optimal frequencies suitable for signalling and know where to ``look'' in frequency space. If we suppose that the signals are diffraction-limited beams and that our search is undertaken over a timespan $t$, we have
\begin{equation}\label{epsd}
  \varepsilon_d = \Omega_\mathrm{r} \cdot \Omega_\mathrm{t} \cdot \frac{t}{\tau},
\end{equation}
where $\Omega_\mathrm{r}$ and $\Omega_\mathrm{t}$ are the solid angles covered by the receiving and transmitting telescopes during the survey, and $\tau$ is the integration time required to achieve the minimum desired signal-to-noise ratio (SNR). Note that $\Omega_\mathrm{t} = {\theta^2}/{4\pi}$ with $\theta = \lambda/D$, where $\lambda$ is the wavelength and $D$ is the size of the transmitting telescope. Clearly, (\ref{epsd}) depends on many parameters, each of which is dictated by the properties of the telescope and mode of functioning. As a result, assigning a point value to this variable is not easy, but this question was investigated in detail for optical SETI by \citet{Lub16}. Two different scenarios corresponding to ``blind'' and ``intelligent'' targeting were considered, and the survey parameters were chosen to be commensurate with current human technology. From Section 3.1 and Figure 4 of \citet{Lub16}, it can be seen that $\varepsilon_d \sim 10^{-6}$ for a distance of $\sim 10$ kpc in the blind detection case, whereas intelligent targeting yields $\varepsilon_d \sim 1$. It should also be noted that the putative existence of isotropic beacons, which would require enormous amounts of power, would also lead to a considerable enhancement of $\varepsilon_d$.

A point worth noting is that $\varepsilon_d \propto (\lambda/D)^2$, implying that its value for radio SETI may be higher compared to optical SETI. Henceforth, we will adopt a fiducial normalization of $10^{-6}$ for $\varepsilon_d$, although its actual value could easily be higher or lower. Lastly, we are confronted with the factor $f_{cs}$ in (\ref{probt}). While the pros and cons of active SETI have been extensively debated \citep{Mus12,Sho13,Bu16,Vak16}, it is worth recalling that the Arecibo telescope has already been used to make a broadcast in 1974. We will work with an optimistic fiducial estimate of unity for $f_{cs}$, but a smaller value cannot be ruled out; see \citet{Sho15} for a related discussion.

\section{The relative likelihood and its implications}\label{SecDel}
We are now in a position to estimate the relative likelihood $\delta$ from (\ref{delta}) by utilizing our preceding discussion. Thus, our final expression is given by
\begin{equation}\label{deltafin}
\delta \sim 0.02\,\alpha\,f_{cs}\,\left(\frac{f_i}{0.1}\right)\left(\frac{L}{10^4\,\mathrm{yr}}\right)\left(\frac{\varepsilon_d}{10^{-6}}\right)\left(\frac{f_d}{0.1}\right)^{-1}.   
\end{equation}
One of the major advantages inherent in the above equation is that both $f_e$ and $f_l$ are absent. The latter in particular is daunting to estimate since the origin of life on Earth and other exoplanets, as well as its characteristic timescale(s), remains one of the fundamental unresolved questions in science \citep{Fry00,Kno15,Lui16}. However, in lieu of these two factors, we have acquired two other factors ($\varepsilon_d$ and $f_d$) but these variables are arguably easier to model for given technologies.

It bears repeating that one of the greatest advantages associated with (\ref{deltafin}) is that the necessity for determining $f_l$ has been eliminated. However, in doing so, we have sidestepped the issue of whether searches for biosignatures or technosignatures should be carried out in \emph{absolute} terms, as opposed to measuring their relative merits. Let us consider an explicit example for the purpose of illustrating our point. In (\ref{Ncalb}), let us suppose that $f_e \sim 1$, i.e. every star has a habitable planet around it - this is very optimistic since it presupposes that a large fraction of all exoplanets are habitable. Thus, using the value of $N_\mathrm{sur}$ from Sec. \ref{SSecBio}, it can be seen that whenever $f_l \lesssim 10^{-5}$, we have $\mathcal{N}_b \lesssim 1$. In the same spirit, $\mathcal{N}_t$ is also likely to be smaller than unity when $f_i$ is sufficiently small, even when optimistic values are specified for the other parameters in (\ref{Ncalta}). 

Thus, the key thing to appreciate is that both $\mathcal{N}_b$ and $\mathcal{N}_t$ become infinitesimally small in the limit $f_l \rightarrow 0$, i.e. if abiogenesis is an extremely rare phenomenon. As the likelihood of abiogenesis does not constitute the subject of our paper, we will not tackle it here,\footnote{The reader may consult \citet{Fry00,ST12,Lui16,SB17} for reviews and analyses of this subject using a wide array of methodologies.} but a couple of general qualitative observations are in order. First, if the origin of life is the real ``bottleneck'' and we end up with both $N_b \ll 1$ and $N_t \ll 1$, given by (\ref{Nbd}) and (\ref{Ntd}) respectively, then the searches for both primitive and intelligent life would have a low chance of success. On the other hand, if the real bottleneck is the emergence of technological intelligence, the search for biosignatures would be heavily favored (by many orders of magnitude) relative to technosignatures. Finally, if $f_i$ is not very small, biosignatures would still merit a higher priority compared to technosignatures, but the latter cannot be wholly set aside in this scenario.

Returning to (\ref{deltafin}), an obvious point that becomes evident upon inspecting this formula is that $\delta \rightarrow 0$ whenever any of the quantities in the numerator approach zero. Let us consider the role of $f_i$ in particular. Many theoretical models posit that the evolutionary history of the Earth was characterized by a small ($<10$) number of major evolutionary transitions \citep{JMS95,KB00,deDu,La09,Sza15}, the most recent of which was the origin of \emph{Homo sapiens} with its advanced tool-making and linguistic capabilities. If each of these transitions was a critical step with a very low probability of occurrence, then it is conceivable that $f_i$ could become extremely small. This argument has been commonly invoked by many evolutionary biologists \citep{Gay64,Mon71,Mayr85}. 

For example, recent mathematical models indicate that the total number of critical steps after the origin of life on Earth leading to the emergence of technological intelligence (humans) is most likely to have been four \citep{Wat08,ML10,LiLo18}. If each critical step had an equal probability on the order of $0.01$ \citep{Cart83}, the cumulative probability for attaining technological intelligence would be $\sim 10^{-8}$. Given that there are $\sim 10^7$ eukaryotic species existing on our planet \citep{MTASW}, of which only one exhibits a high degree of technology, this lends some credence to the idea that $f_i \ll 1$ is possible. An important point to be noted, however, is that the critical step model is only one possible theoretical model. Instead, it is conceivable that the major evolutionary innovations on Earth and elsewhere arose via the ``many paths model'', implying that the evolution of complex, animal-like life may be nearly inevitable provided that life has originated and sufficient energy fluxes are available \citep{BaSM16,SB17}. In this event, as the likelihood of complex life becomes high (given abiogenesis), this may lend some credence to the notion that $f_i$ is not extremely small. 

Apart from this option, several other arguments have been presented in favor of $f_i \sim 1$, and we refer the reader to \citet{Mar15} for a succinct summary, whereas a critique of some of the traditional arguments invoked in favour of a high value of $f_i$ can be found in \citet{Lin09}.  Most of the conventional arguments rely on the fact that a large number of ``human'' traits - such as  ``high'' intelligence, culture, tool-making, theory of mind, and symbolic communication - have been documented, albeit controversially, to varying degrees in certain animals \citep{RD05,WR14,DW16}.\footnote{Charles Darwin espoused a similar standpoint in \emph{The Descent of Man} \citep{Dar}, where he wrote: ``Nevertheless the difference in mind between man and the higher animals, great as it is, certainly is one of degree and not of kind.''} Some proponents of SETI have also invoked the apparent trends in biological complexity and brain size, which tend to be approximately characterized by exponential or power-law growth \citep{Jer73,Russ83,Ros13}, to argue that high technological intelligence is comparatively likely once abiogenesis has occurred on a planet. Recent breakthroughs in studies of evolutionary convergence have also been offered as evidence in favor of a relatively high probability for the emergence of humanoids provided that abiogenesis did occur successfully \citep{DD02,Mor03,FM14}. In this context, the recent work by \citet{Cir18} also presents several arguments for $f_i \sim 1$ based on evolutionary convergence, and critiques some of the common arguments offered in favor of $f_i \ll 1$. 

Next, let us turn our attention to $L$ and $\varepsilon_d$ because there is a common theme that runs through both of them. We will focus on $L$ first, as the same argument can be repeated for $\varepsilon_d$. Consider the example where $1\%$ of all technological species have a long lifetime of $\sim 10^8$ yr, whereas the rest of them have a short lifetime of $< 10^4$ yr. Upon calculating the average lifetime, it can be seen that $L \sim 10^6$ yr in this case, i.e. there can be situations in which the tail of the lifetime probability distribution function dominates the average. This is a point that has been appreciated since the 1960s \citep{SS66}, and it offers an avenue by which the value of $L$ may be increased in (\ref{deltafin}). Along these lines, even if a small fraction of technological species opt for isotropic (not beamed) signals despite the power requirements, the average value of $\varepsilon_d$ can be increased by a few orders of magnitude. On the other hand, if all technological species are short-lived - one of the explanations commonly offered for Fermi's paradox during the Cold War - the magnitude of $\delta$ will be correspondingly lowered. 

This leaves us with the factors $f_{cs}$, $f_d$ and $\alpha$. The former is dependent upon sociological considerations that are hard to deduce \emph{a priori}. However, if the benefits of communicating with, or merely seeking, other technological species outweigh the risks \citep{Loeb}, it seems plausible to us that $f_{cs}$ should be close to unity once a particular species has attained a certain level of stability. Turning our attention to $f_d$, it seems unlikely that $f_d \ll 1$ despite the many false positives/negatives possible for exoplanets in the HZ of M-dwarfs, especially since new strategies for distinguishing between false and real positives are being formulated \citep{CKT18}. 

Lastly, note that $\alpha$ represents the number of ``outposts'' that a technological species has per star. For our specific example, each ``outpost'' must also be capable of sending targeted optical/radio signals detectable throughout the Galaxy. As before, it should be noted that $\alpha$ refers to the \emph{average} value per species. Hence, even if a small fraction of highly advanced species have sent a large number of probes and settled multiple sites, the value of $\alpha$ may become much higher than unity. However, this would bring it into some conflict with the so-called Fermi's paradox, since we would have to explain why none of these probes (or their signals) have been detected thus far \citep{Brin,Webb}. Nevertheless, since we have not comprehensively surveyed the Solar neighbourhood, the presence of such probes cannot be ruled out definitively \citep{Frei83,LT12,HMK12}.\footnote{In fact, we cannot even conclusively rule out the presence of prior technological species on the Earth if it was alive millions of years ago \citep{Dav12,Wri18,SF18}.} In the future, interstellar objects passing through the Solar system, like `Oumuamua, can be searched for signs of electromagnetic signals \citep{ES18,TK18} or be subjected to detailed \emph{in situ} exploration \citep{SL18}.

It should also be noted that $\alpha$ is expected to be correlated with $L$ under certain circumstances. Suppose that the rate of settling a new site, or sending out a new spaceship capable of Galaxy-spanning communication, is constant and denoted by $\Lambda$. In this case, the maximum number of sites settled by a typical technological species is given by $\alpha = 1 + \Lambda L$, where $\alpha \rightarrow 1$ when $L \rightarrow 0$ since the multiplicity factor has a lower bound of unity. While the exact value of $\Lambda$ remains unknown, a fairly conservative value would be $\Lambda \sim 10^{-4}$/yr \citep{Jon81,WM14}. For this choice of $\Lambda$, we arrive at
\begin{equation}\label{alpha}
    \alpha \sim 1 + \left(\frac{L}{10^4\,\mathrm{yr}}\right).
\end{equation}

Although we have hitherto retained all possible factors, based on our prior arguments, it seems to us that the maximum variability is encoded in $\alpha$, $f_i$ and $L$ with respect to the other parameters in (\ref{deltafin}). Hence, upon using (\ref{alpha}), we propose that the following formula can be viewed as a simplified version of (\ref{deltafin}) for the relative likelihood $\delta$ that retains the essential features:
\begin{equation}\label{deltasim}
\delta \sim 0.02\,\left(\frac{f_i}{0.1}\right)\left(\frac{L}{10^4\,\mathrm{yr}}\right)\left[1 + \left(\frac{L}{10^4\,\mathrm{yr}}\right)\right].   
\end{equation}

\section{Alternative searches for technosignatures and the relative likelihood of detection}\label{SecASETI}
Hirtherto, we have restricted ourselves to electromagnetic SETI, but the importance of artifact SETI, i.e. detecting artifacts of technological species, has gained wider appreciation recently \citep{BCD11,Car12,WM14}. At this stage, it is necessary to differentiate between searching for artifacts that can be built by human-level technology, and those that can be built by more advanced species. The latter includes megastructures and macro-engineering projects that we shall discuss later. Before proceeding further, we observe that the proposed distinction between intentional electromagnetic signals, artifacts at the level of human technology, and megastructures ought not be regarded as being clear-cut. It is quite conceivable that artifacts, especially megastructures, are already present in our catalogs but have not been identified yet; one such example has been discussed by \citet{Cirk16}.

Examples of artifacts and signals that can be produced by human-level technology include photovoltaic arrays for utilizing stellar energy \citep{LL17}, artificial lights from cities \citep{LT12}, global warming \citep{KB15}, industrial pollution via chlorofluorocarbons in the atmosphere \citep{LGL14}, and geostationary/geosynchronous satellites in orbit \citep{SN18}. Since all of these technosignatures are intimately tied to detecting and characterizing exoplanets, the corresponding values for $\mathcal{N}_t$ and $\mathcal{P}_t$ will closely resemble (\ref{Ncalb}) and (\ref{Probb}).

Since we are interested only in detection using present-day technology, we will focus on characterizing transiting planets using the JWST. In this case, $\mathcal{N}_t$ can be expressed as
\begin{equation}\label{Ncaltb}
    \mathcal{N}_t = N_\mathrm{sur} \cdot f_e \cdot f_l \cdot f_i \cdot f_a,
\end{equation}
where $f_a$ is the fraction of the stellar lifetime that these artifacts are detectable. Upon comparing the above equation with (\ref{Ncalta}), two major differences exist. First, since we have assumed that the species under consideration possess a level of technology commensurate with that of humans, we have set the multiplicity factor to be roughly unity. Second, in place of $L/t_\star$, we have introduced the factor $f_a$. This is because of the fact that these species do not need to be currently alive in order for some of their artifacts to still be detectable. In other words, we contend that $f_a \gtrsim L/t_\star$ and several technosignatures of extinct species have been argued to persist on timescales of $\sim 10^5$ yrs \citep{SFJ16}, although megastructures might last even longer. The longevity of surface-based artifacts will be controlled by a wide array of geological processes - for example: erosion by winds and flowing water, volcanism, large-scale glaciation and plate tectonics - and all traces of technology may be eliminated over a timescale of $\gtrsim 10^6$ yrs \citep{Wri18,SF18}. On account of the above reasons, we will adopt a somewhat conservative fiducial normalization of $10^{-5}$ for $f_a$. 

Next, let us turn our attention to $\mathcal{P}_t$, which is given by
\begin{equation}\label{Probt}
    \mathcal{P}_t = f_q \cdot f_t,
\end{equation}
and therefore closely resembles (\ref{Probb}). We have still retained the factor of $f_q$ since frequent flares and superflares are not expected to be conducive to complex surface-based life \citep{SW10,ML17}. The chief difference between (\ref{Probt}) and (\ref{Probb}) is the latter contains the extra factor of $f_d$ to account for the possibility of false positives/negatives insofar biosignature gases are concerned. One of the basic premises underlying the search for technosignatures is that the likelihood of false positives is much lower,\footnote{The most unambiguous sign of life on Earth was through the detection of narrow-band radio signals by the Galileo spacecraft during a flyby mission \citep{ST93}.} as a result of which we have dropped the equivalent of $f_d$ for technosignatures. If necessary, this factor can be easily included within our formalism.

Upon using the above equations, the relative likelihood turns out to be
\begin{equation}\label{deltaart}
    \delta \sim 10^{-5}\,\left(\frac{f_i}{0.1}\right)\left(\frac{f_a}{10^{-5}}\right)\left(\frac{f_d}{0.1}\right)^{-1}.
\end{equation}
When we compare (\ref{deltaart}) against (\ref{deltasim}), it can be seen that the latter is higher than the former by about three orders of magnitude. It should be reiterated here that there are several uncertainties in both formulae, but the comparison of these two equations suggests that searching for electromagnetic signals might be a more productive strategy compared to the search for artifacts. The primary reason behind this result stems from the fact that the potential search volume for electromagnetic signals is much larger, whereas the detection of planet-based artifacts is reliant on using transmission/eclipse spectroscopy or direct-imaging in the future.

Lastly, let us consider the possibility of searching for evidence of macro-engineering and megastructures. The best-known example in this category is the Stapledon-Dyson sphere \citep{Stap37,Dys60} that encompasses its host star to collect a significant fraction of the stellar energy. Other examples of planet-sized megastructures detectable through light curves include Shkadov thrusters \citep{For13}, mirrors \citep{KSL15}, starshades \citep{Gai17}, and other artificial objects \citep{Arn05,WC16}. Many of these macroengineering projects fall within the capacity of species that have attained the Kardashev II class \citep{Kar64}. Looking further beyond, the signatures of species belonging to the Kardashev III class are expected to be manifested on a galactic scale. Searches for Kardashev III species have looked for alien waste heat \citep{WG14}, anomalies in the optical Tully-Fisher relationship \citep{ZCA15} and missing stars \citep{VIB16}. The key thing to note in dealing with the Kardashev II and III classes is that the accessible survey volume happens to be very large. For instances, the Gaia mission is expected to enable the search for Stapledon-Dyson spheres over a significant proportion of the Galactic stellar population \citep{ZK18}. The search for Kardashev III species encompasses an even higher number of stars - for example, the $\hat{\mathrm{G}}$ infrared search spanned $\sim 10^5$ galaxies \citep{GW15}. 

Thus, if we attempt to calculate $\mathcal{N}_t$ by means of (\ref{Ncalta}), the factor $N_g$ should be replaced by $\sim 10^5 N_g$. Since $\delta \propto \mathcal{N}_t$, this would appear to naively suggest that the characteristic value of $\delta$ will be raised by $5$ orders of magnitude. However, this analysis neglects a crucial point - recall that $L$ represents the average lifetime of the species such that it will be detectable. Hence, in this particular case, $L$ denotes the average lifetime of a species at the Kardashev III level. Thus, there is a possibility that the average lifetime for Kardashev III species is very short because of self-annihilation and other reasons. This can happen because technology need not progress along a monotonically increasing path \citep{Den11}, and advanced species might either self-destruct or opt for sustainability over monotonic expansion \citep{Webb,FCAK}. Since we have no knowledge whatsoever as to what fraction of technological species (at the level of humans) can eventually attain a Kardashev III stage, we will not attempt to quantify the value of $\delta$ for such surveys. 

\section{The implications for funding different search strategies}\label{SecFun}
Hirtherto, we have concerned ourselves only with analyzing $\delta$, but we shall now consider the implications for funding. In this context, we shall focus only on the USA herein for simplicity. 

The JWST, which can be used for studying biosignatures as discussed in Sec. \ref{SSecBio}, is being built and launched at a total cost of $\sim \$10$ billion and has an operational lifetime of $\lesssim 10$ yrs.\footnote{\url{https://www.jwst.nasa.gov/facts.html}} Hence, the corresponding funding level assigned to detecting life via exoplanetary biosignatures is $\lesssim \$1$ billion/yr; note that the actual value will be lower by a factor of a few since the JWST is expected to undertake other tasks as well. This estimate is roughly comparable to the total astrobiology budget of $\sim \$1$ billion/yr estimated in \citet{KGO18}.\footnote{This amount ostensibly includes the funding for space missions and not just the funding for the NASA exobiology program and the NASA Astrobiology Institute.} Therefore, we specify the annual budget for detecting biosignatures over the next decade to be of order $\$0.5$ billion/yr in our model.

If more worlds with life can be detected by means of biosignatures as opposed to technosignatures, it seems reasonable to conclude that a lower amount of funding can be devoted to the latter compared to the former. We will use the simple ansatz wherein the amount of funding that deserves to be allocated is proportional to the theoretical number of worlds that can be detected using a particular method. For this case, denoting the annual funding for technosignatures by $C$, we obtain $C = \$\,(\delta/2)$ billion/yr. Previously, we found that the maximum value of $\delta$ was attained for electromagnetic SETI in Sec. \ref{SecDel}. We can make use of either (\ref{deltafin}) or (\ref{deltasim}),  but we shall use the latter because it retains the most essential parameters $f_i$ and $L$ and possesses a simpler form. Thus, the value of $C$ becomes
\begin{equation}\label{Csim}
C \sim \$10\,\mathrm{million/yr}\,\left(\frac{f_i}{0.1}\right)\left(\frac{L}{10^4\,\mathrm{yr}}\right)\left[1 + \left(\frac{L}{10^4\,\mathrm{yr}}\right)\right].   
\end{equation}
From the above formula, two aspects stand out. First, if at least one of the factors $f_i$ and $L$ becomes very small, the amount of funding that is justifiable in searching for technosignatures also drops steeply. Second, if we assume that the characteristic values chosen for $f_i$ and $L$ are indeed valid, a total funding of $\$100$ million per decade is justified for detecting technosignatures. This total coincides with the funding that has been recently allocated for the Breakthrough Listen initiative.

However, there is a subtle point that is overlooked in many analyses of this kind. Consider the following situation: there are $10$ worlds with microbial life that were detected through biosignature gases, and $10$ worlds with technological life found through searches for technosignatures. Are the two cases fully equivalent? If we were merely interested in finding extraterrestrial life, regardless of its complexity, the two outcomes could be treated on an equal footing. Yet, there is a distinction because the \emph{impact} of detecting technological intelligence is not the same as that of finding microbes \citep{Tart07,KGO18}. Evidently, there would be an anthropocentric bias since we are probably more predisposed to rank technological intelligence higher because we categorize these traits (rightly or wrongly) as ``human''. Hence, finding evidence of technological intelligence elsewhere ought to have a greater impact with respect to detecting microbial life. 

Another key point worth highlighting here is that detecting extraterrestrial intelligence (ETI) would, in all likelihood, \emph{raise} the prospects for finding microbial life as well. This is because, based on the course of evolution on Earth, it seems reasonable to assume that any technological advanced species must have had a very large number of non-technological precursors (especially microbes). Thus, finding evidence of technological intelligence lends credibility to the notion that very large numbers of primitive extraterrestrial organisms exist, thereby improving the chances for detecting them via searches for biosignatures.

The impacts of detecting ETI are many \citep{Fin90,Kat11,DZ11,Cap13,Dick}, and include a wide range of potentially positive benefits, such as gaining new scientific and technological knowledge \citep{BHD11}. Even if this detection did lead to negative outcomes, it does not alter the notion that the overall impact is likely to be considerable. On account of these reasons, we suggest that the right-hand-side of (\ref{Csim}) should be multiplied by another factor ($\kappa$) that quantifies the impact of detecting technological intelligence relative to microbial life. Consequently, we are confronted with the question of the typical value(s) of $\kappa$. At the very least, it seems reasonable to conclude that it should be a factor of $\mathcal{O}(10)$ because of the greater impact on human culture, science and religion. In this event, $C$ is expressible as
\begin{equation}\label{Ckap}
C \sim \$100\,\mathrm{million/yr}\,\left(\frac{\kappa}{10}\right)\left(\frac{f_i}{0.1}\right)\left(\frac{L}{10^4\,\mathrm{yr}}\right)\left[1 + \left(\frac{L}{10^4\,\mathrm{yr}}\right)\right].   
\end{equation}

Let us, however, consider a more unorthodox proposal for $\kappa$. We hypothesize that $\kappa \sim 1/f_i$, i.e. the relative impact is inversely proportional to the rarity of technological intelligence with respect to microbial life. In qualitative terms, this essentially amounts to saying that the weight (or impact) we attach to the detection of technological intelligence increases if the likelihood of ETI in the Universe becomes smaller. A loose analogy can be drawn with the law of downward-sloping demand for commodities with unit elasticity \citep{SN09,KW15}, wherein the price and the quantity are inversely proportional to each other. If we pursue this line of reasoning, the ``quantity'' can be associated with $f_i$ (relative number of ETIs) and the ``price'' with $\kappa$ (impact). When we adopt this formulation for $\kappa$, the funding is given by
\begin{equation}\label{Ccom}
C \sim \$100\,\mathrm{million/yr}\,\left(\frac{L}{10^4\,\mathrm{yr}}\right)\left[1 + \left(\frac{L}{10^4\,\mathrm{yr}}\right)\right].   
\end{equation}
As a result of our ansatz, we see that $f_i$ has dropped out from the formula, which is helpful since this is arguably the parameter subject to the most uncertainty. If the characteristic value of $L$ is valid, this formula predicts that an expenditure of $\$1$ billion per decade would be justified - an amount that is about an order of magnitude higher than the current spending. However, if the value of $L$ is significantly lowered, note that the value of $C$ will also be reduced by the same factor. If we choose a minimum value of $L \sim 10^3$ yr, which might be plausible based on our own civilization \citep{Ma07}, we obtain $C \sim \$10$ million/yr after using (\ref{Ccom}). This is a fairly stable result since (\ref{Ccom}) depends only on $L$ and not on $f_i$. If we use the same lower bound for $L$ in (\ref{Ckap}), we end up with $C \sim \$10$ million/yr once again, but only if the restrictive criterion $\kappa f_i \sim 1$ is assumed to be valid.

\section{Conclusion}\label{SecConc}
In this paper, we assessed the relative likelihood of detecting primitive and intelligent life, by means of biosignatures and technosignatures respectively, using state-of-the-art telescopes. We followed this up by addressing the ensuing implications with regards to funding life-detection searches for biosignatures and technosignatures.

We focused on searches that are accessible by current technology: the detection of biologically produced gases by means of transmission spectroscopy using a telescope like the JWST and the detection of technosignatures through optical or radio signals of an artificial origin. While our final result (\ref{deltafin}) does have a number of unknown variables, we estimated that the likelihood of detecting intelligent life might be two orders of magnitude smaller compared to the detection of primitive life. It should be noted that this estimate can be seen as an upper bound since it will be lowered significantly if technological intelligence is rare or short-lived. We also considered the possibility of looking for artifacts instead of electromagnetic signals, and concluded that the latter is still likely to be more effective because of its greater search volume. 

Subsequently, we discussed the implications for funding the search for primitive and intelligent life. As primitive life has a much higher likelihood of being detected, we concluded that the majority of funding should be allocated to the search for biosignatures. The discussion in Sec. \ref{SecFun}, especially following (\ref{Ccom}), suggests that SETI might merit a minimum federal funding of $\$10$ million/yr provided that the average lifetime for technological species is $\sim 10^3$ yrs. This estimate matches the amount of funding that has been recently allocated to SETI by the Breakthrough Listen initiative, and the bill that is being considered by the USA House of Representatives.

In our analysis, we adopted an approach along the lines of the classical Drake equation. Hence, many of the critiques levelled against the Drake equation are also applicable here \citep{Dick15}. For instance, our final results depend on factors like the detectable lifetime of a technological species that are poorly understood and therefore subject to much uncertainty \citep{SS66}. A more realistic treatment should model the various factors based on a suitable statistical framework \citep{Mac11,GBB11}. Moreover, our results do not account for spatial and temporal variations since all of the factors are implicitly treated as being constant \citep{Bur06}. Despite these caveats, our approach may provide a heuristic framework for facilitating further discussion regarding the relative merits of various search strategies and the amount of funding that ought to be allocated to them. 

\section*{Acknowledgments}
We thank Michael Hippke, Jill Tarter, and the two reviewers for their constructive feedback regarding the paper. This work was supported in part by the Breakthrough Prize Foundation for the Starshot Initiative, Harvard University's Faculty of Arts and Sciences, and the Institute for Theory and Computation (ITC) at Harvard University.


\end{document}